\documentclass{iaus}
\usepackage{graphicx}

\title[Stellar populations of NGC 7582] %% give here short title %%
{The stellar populations of the \\ AGN/Starburst galaxy NGC7582}

\author[T.V. Ricci et al.]   %% give here short author list %%
{T.V. Ricci$^{1\dagger}$, J.E. Steiner$^1$, R.B. Menezes$^1$, A. Garcia-Rissmann$^2$ \and R. Cid Fernandes$^3$}

\affiliation{$^1$Instituto de Astronomia, Geof\'isica e Ci\^encias Atmosf\'ericas, Universidade de S\~ao Paulo, \\
Rua do Mat\~ao, 1226, S\~ao Paulo - SP, Brasil \\[\affilskip]
$^2$Gemini Observatory \\ [\affilskip] $^3$Universidade Federal de Santa Catarina \\ $^{\dagger}${\tt tiago@astro.iag.usp.br}}

\pubyear{2009}
\volume{262}  %% insert here IAU Symposium No.
\pagerange{1--2}
\jname{Stellar Populations - Planning for the Next Decade}
\editors{Gustavo R. Bruzual \& Stephane Charlot}
\begin{document}

\maketitle

\begin{abstract}

NGC 7582 is defined as a Starburst/AGN galaxy, since its optical and X-Ray spectra reveal both characteristics. In this work, we show the results of a stellar population modeling in a datacube taken with the Gemini South telescope. We found that $\sim$ 90\% of the light in the field of view is emitted by stars that are less than 1 billion years old. A strong burst occurred about $\sim$ 6 million years ago and has nearly solar metallicity. We also found a Wolf-Rayet cluster.
\end{abstract}

\section{Introduction}

Starburst galaxies are characterized by their strong and narrow emission lines in the
optical spectra, due to a strong star formation in the galaxy center. Seyfert galaxies have
broader emission lines and are interpreted as being produced by photoionization by a power-law continuum associated to a supermassive black hole. NGC 7582 has a typical H II region spectrum in the optical \cite[(Veron et al. 1981)]{Veron et 
al.(1981)} but its X-Ray emission is typical of a Seyfert 1 galaxy \cite[(Ward et al. 1978)]{Ward et al.(1978)}. In this work our goal is to analyze the stellar population located in the 3.5'' x 5'' central region of NGC 7582.

\section{Observations and Methodology}

Our observations were taken with the GMOS-IFU \cite[(Allington-Smith et al. 2002)]{Allington-Smith et al.(2002)}  on the Gemini South telescope. We used the B600-G5323 grating, producing a spectral range of 4230-7070 $\dot{A}$ with R = 2400 (as measured from the 5577 $\dot{A}$ sky line). The final datacubes are corrected from the atmospheric differential refraction; we also performed a Richardson-Lucy deconvolution with 6 iterations on the image in each spectral pixel. The stellar population synthesis was made using the {\sc starlight}  code \cite[(Cid Fernandes et al. 2005)]{Cid Fernandes et al. (2005)} on the spectrum of each spatial pixel, with a sampling of 0.05'' x 0.05''. Maps of different population were constructed, revealing age and metallicity structures. These results will be published elsewhere.

\section{Results and conclusions}

\begin{figure}[t]
\begin{center}
%\setcaptionmargin{1cm}
\begin{minipage}[b]{0.45\linewidth}
\includegraphics[width=1.0 \columnwidth,angle=0]{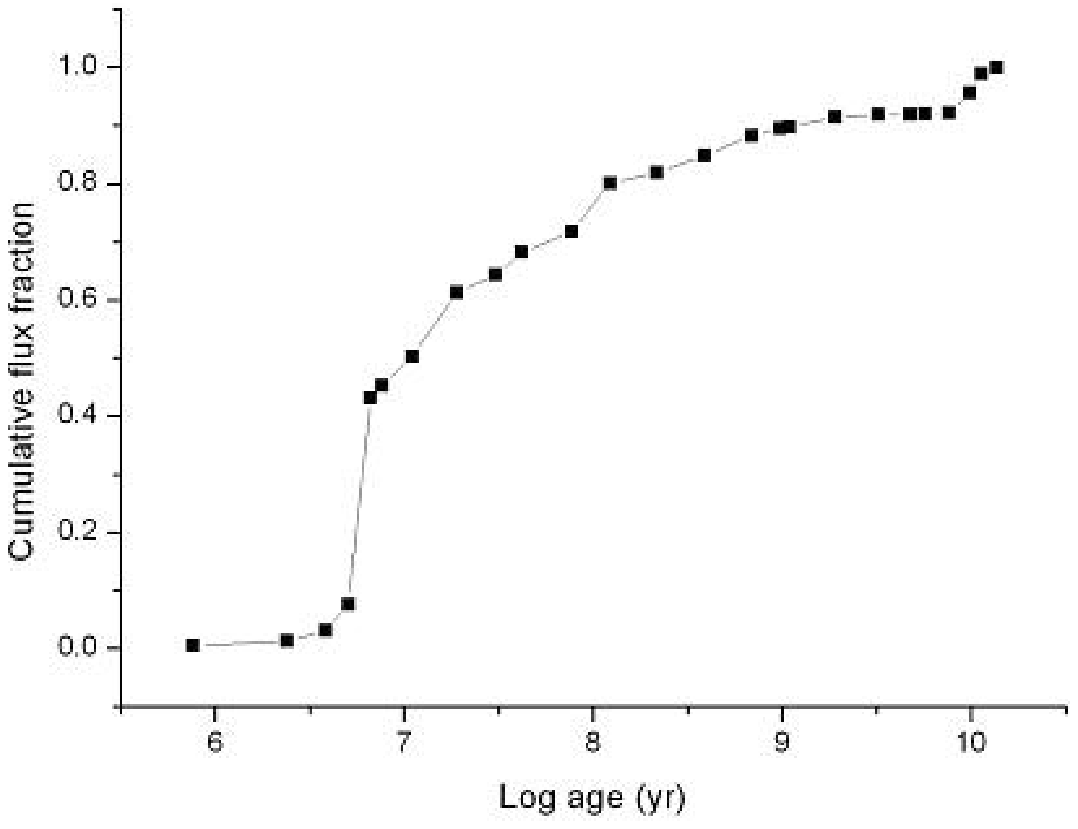}
\caption{Cumulative flux fraction as a function of the stellar age of the entire field of view.}
\label{fig1}
\end{minipage} \hfill
\begin{minipage}[b]{0.45\linewidth}
\includegraphics[width=1.0 \columnwidth,angle=0]{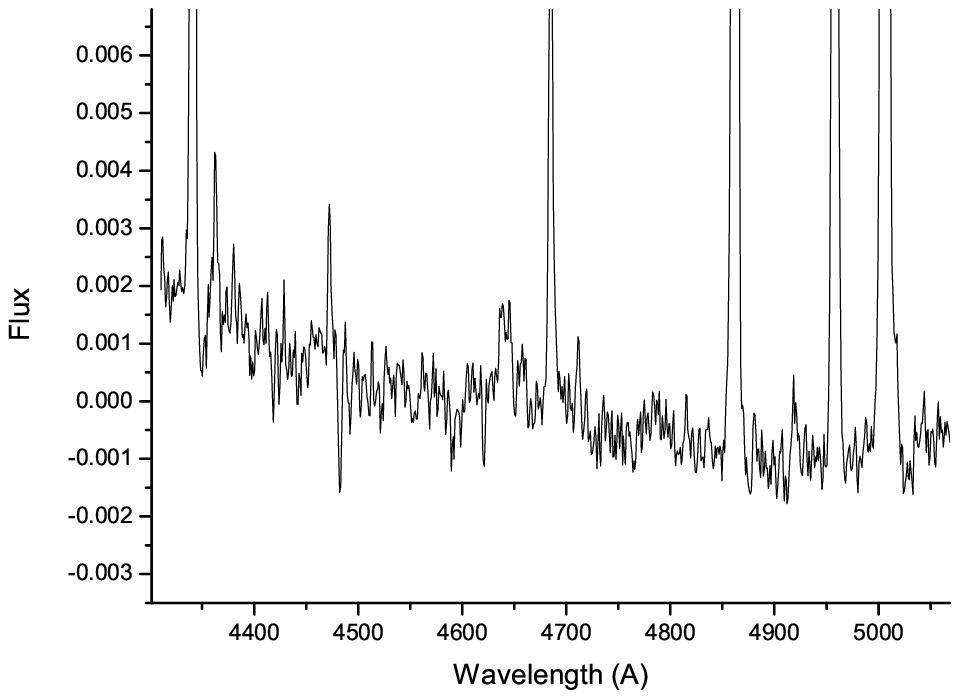}
\caption{The spectrum showing the WR feature at 4600 to 4700 $\dot{A}$. This spectrum was extracted  at a position 1.4" west and 1.0" north of the AGN.}
\label{fig2}
\end{minipage}
\end{center}
\end{figure}

Figure \ref{fig1} shows the cumulative flux fraction as a function of stellar age. It is clear that about 90\% of the light in the field of view is emitted by stars that are less than 1 billion years old. There seems that a more or less continuum star formation episodes have occurred in this period. Nearly half of the light is emitted by stars less than 10 million years old. A strong event occurred $\sim$ 6 million years ago. This young stellar population shows nearly solar metallicity and is concentrated in three clumps that might be associated with the ionizing sources of the three observed H II regions. One of the clumps, located nearly 1.4'' West and 1.0'' North of the AGN presents a Wolf-Rayet feature, shown in Figure \ref{fig2}. This feature has broad C III/N III + He II emission between 4600 $\dot{A}$ and 4700 $\dot{A}$.
\\
\\
I would like to thank FAPESP for the financial support.

\end{document}